# Intersatellite-link demonstration mission between CubeSOTA (LEO CubeSat) and ETS9-HICALI (GEO satellite)

Alberto Carrasco-Casado[1], Phong Xuan Do[2], Dimitar Kolev[1], Takayuki Hosonuma[2], Koichi Shiratama[1], Hiroo Kunimori[1], Phuc V. Trinh[1], Yuma Abe[1], Shinichi Nakasuka[2], and Morio Toyoshima[1]

[1] Space Communications Laboratory
National Institute of Information and Communications Technology (NICT), Japan

[2] Department of Aeronautics and Astronautics
The University of Tokyo, Japan

*Abstract*— LEO-to-GEO intersatellite links using laser communications bring important benefits to greatly enhance applications such as downloading big amounts of data from LEO satellites by using the GEO satellite as a relay. By using this strategy, the total availability of the LEO satellite increases from less than 1% if the data is downloaded directly to the ground up to about 60% if the data is relayed through GEO. The main drawback of using a GEO relay is that link budget is much more difficult to close due to the much larger distance. However, this can be partially compensated by transmitting at a lower data rate, and still benefiting from the much-higher link availability when compared to LEO-to-ground downlinks, which additionally are more limited by the clouds than the relay option. After carrying out a feasibility study, NICT and the University of Tokyo started preparing a mission to demonstrate the technologies needed to perform these challenging lasercom links. Furthermore, to demonstrate the feasibility of this technique, an extremely-small satellite, i.e. a 6U CubeSat, will be used to achieve data rates as high as 10 Gbit/s between LEO and GEO. Some of the biggest challenges of this mission are the extremely low size, weight and power available in the CubeSat, the accurate pointing precision required for the lasercom link, and the difficulties of closing the link at such a high speed as 10 Gbit/s.

## I. INTRODUCTION

NICT has a long experience in space laser communications, starting with the ETS-VI (Engineering Test Satellite VI) satellite, which demonstrated lasercom downlinks with the NICT 1-m optical ground station (OGS) back in 1994 from the GEO orbit for the first time [1]. Focusing on low-Earth orbit (LEO) satellites only (Fig. 1), NICT carried out the first successful bidirectional intersatellite and LEO-ground lasercom demonstration with OICETS (Optical Interorbit Communications Engineering Test Satellite), launched in 2005, achieving 50 Mbit/s [2] [3]. Less than 10 years later, SOTA (Small Optical TrAnsponder) was the first lasercom terminal onboard a microsatellite [4]. This microsatellite was called SOCRATES (Space Optical Communications Research Advanced Technology Satellite) and it was used to carry out a variety of different experiments starting in 2014, including up-to-10-Mbit/s downlinks and space-QKD experiments [5]. In this paper, the concept of a new NICT's lasercom terminal onboard a LEO nanosatellite, planned to be launched around 2023, is discussed.

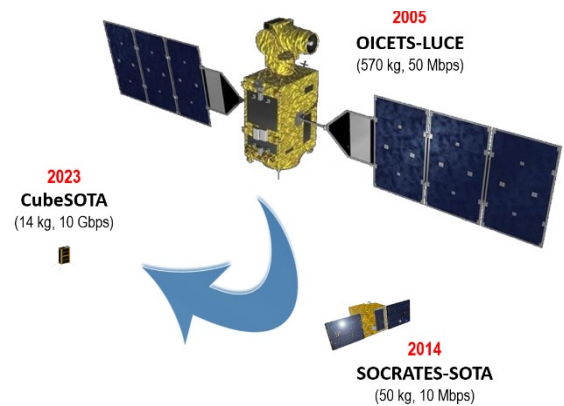

Fig. 1. History of NICT's lasercom LEO satellites (to scale).

Fig. 2 shows the evolution of satellite mass and lasercom bitrate achieved by the NICT's satellites launched to a LEO orbit introduced in the previous paragraph and shown in Fig. 1. The mass of the satellite has experienced a notable reduction of more than one order of magnitude while the communication bitrate has increased by three orders of magnitude in a time span of less than 20 years.

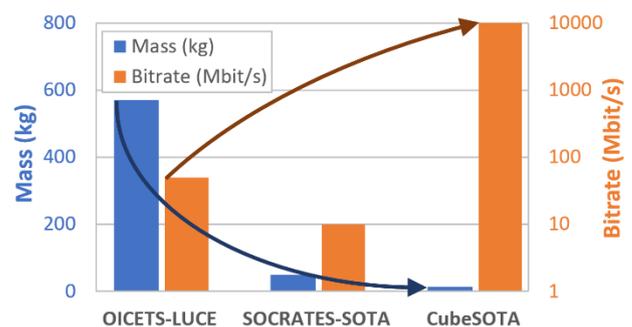

Fig. 2. Evolution of NICT's lasercom LEO satellites.

Among some of the technologies that have enabled this impressive progress are the microsatellite and nanosatellite (i.e. CubeSat) platforms that have democratized many of the space technologies previously reserved only for big and costly satellites. Specifically, the attitude determination and control systems (ADCS) have recently made possible to go without a gimbal mechanism for the coarse pointing, and instead using the body pointing of the satellite, which greatly alleviates the lasercom payload.

## II. LEO-TO-GEO LASERCOM INTERSATELLITE LINKS

LEO-to-GEO intersatellite links using laser communications provide important benefits to greatly enhance applications such as downloading big amounts of data from LEO satellites by using the GEO satellite as a relay whether the downlink from GEO to ground is carried out by RF or lasercom. As shown in the comparison of table 1, by using this strategy, the total availability of the LEO satellite can be increased from less than 1% if the data is downloaded directly to the ground, up to almost 60% if the data is relayed through the GEO relay.

For this simulation, ISS orbit was assumed for LEO and ETS-9 satellite was assumed for GEO. In average, there are about 15 links per day from LEO to GEO with a duration of ~1h, versus about 4 links per day from LEO to ground with a duration of less than 5 minutes. For LEO-GEO, the access was assumed to be limited only by direct line of sight, and in the case of LEO-ground limited by a 10º OGS elevation angle but not by clouds. Additionally, two cases were considered in the case of LEO-ground, where only nighttime links can be performed and where both nighttime and daytime links are allowed.

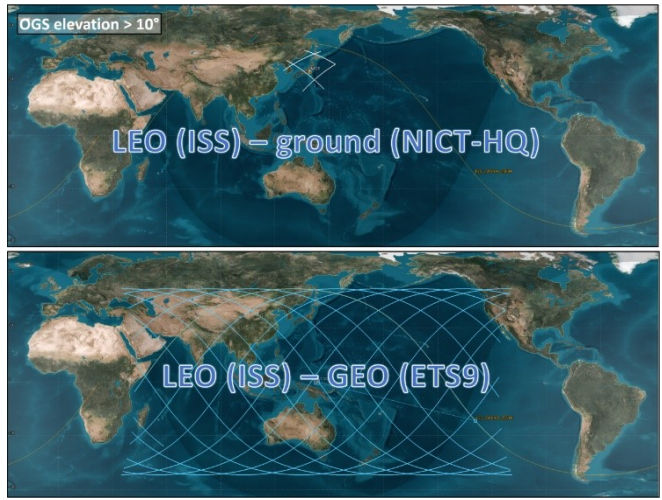

Fig. 3. 1-day LEO-OGS vs LEO-GEO accesses comparison.

Figure 3 shows a comparison between the access time during 1 day by downloading the data directly to the Earth from LEO compared to transmitting the data to the GEO relay.

The main drawback of using a GEO relay is that link budget is much more difficult to close due to the much larger distance, which translates into close-to-30 dB of additional losses in the intersatellite case, although with a smaller variability. However, this can be partially compensated by transmitting at a lower data rate, and still benefiting from the much-higher link availability when compared to LEO-to-ground downlinks, which additionally are more limited by the clouds than the relay option.

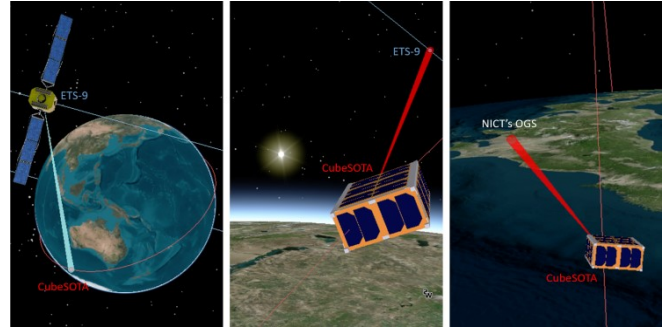

Fig. 4. Conceptual illustration of CubeSOTA experiments.

NICT and the University of Tokyo have started a collaboration to carry out a demonstration mission to test the technologies needed to perform these challenging lasercom links. Furthermore, to demonstrate the feasibility of this technique, an extremely small satellite, i.e. a CubeSat, will be used to achieve data rates as high as 10 Gbit/s. The mission is called CubeSOTA as a continuation of the SOTA achievements. Fig. 4 shows a conceptual illustration of this mission. Some of the biggest challenges are the extremely-low size, weight and power available in the CubeSat, the accurate pointing precision required for the lasercom link, and the difficulties of closing the link at such a high speed as 10 Gbit/s.

A study conducted by the National Academies of Sciences, Engineering, and Medicine on the scientific and technological potential of CubeSats [6] determined that the following four areas will have the largest impact on science missions: high-bandwidth communications, precision attitude control, propulsion, and miniaturized instrument technology. The first of these areas is the main goal of the CubeSOTA demonstration mission, while exploring the limits of the second one is another important objective, as well as advancing in the fourth one, specifically related to the lasercom technologies.

Table 1. LEO-GEO vs LEO-OGS 1-year access summary.

|  | LEO-GEO ☀☾ | LEO-OGS ☀☾ | LEO-OGS ☾ |
|---|---|---|---|
| Duration | 1 Jan 2023 – 1 Jan 2024 | | |
| Location | • LEO = ISS orbit (390 km)<br>• GEO = ETS-9 orbit | • LEO = ISS orbit (390 km)<br>• OGS = NICT OGS (Koganei) | |
| Access type | • Direct line of sight | • No clouds<br>• OGS elevation > 10° | • No clouds<br>• OGS elevation > 10°<br>• Only nighttime links |
| Access time | 5,046 hours | 118 hours | 43 hours |
| Total accesses | 5,332 links (average: 15/day) | 1,463 links (average: 4/day) | 536 links (1/day) |
| Average link | 56.8 min | 4.9 min | |
| Average range | 39,693 km | 1,103 km | |
| Free-space losses | Average = 290 dB (Δ = 1.9 dB) | Average = 259 dB (Δ = 11.4 dB) | |
| Total availability | 57.60% | 1.35% | 0.50% |

## III. LEO CubeSat to GEO ETS-9 demonstration

Table 2 shows the basic specifications of the CubeSOTA mission. A 6U CubeSat platform was selected to accommodate the bus system in ~3U and the lasercom payload in the other ~3U. Its deployment was assumed to be done from the Japanese Experiment Module (JEM) in the International Space Station (ISS) by using the JEM Small Satellite Orbital Deployer (J-SSOD). For this reason, the same ISS orbit was assumed for CubeSOTA. Deploying satellites from the J-SSOD has a few advantages such as the frequent launch opportunities and the more benign environment, with no shock and attenuated vibration because it is delivered with the pressurized cargo to the ISS [7]. The launch is foreseen to be during 2023, approximately one year after ETS-9 launch, with an operation longer than 6 months (total lifetime of a 6U CubeSat launched to a 400-km orbit is ~1.5 years [8]).

Table 2. Basic specifications of CubeSOTA.

| Satellite name | CubeSOTA |
|---|---|
| Satellite bus | The University of Tokyo |
| Payload | NICT's Space Communications Lab. |
| Orbit | LEO-ISS (~390 km) |
| CubeSat format | 6U (340.50×226.30×100.00 mm) |
| Mass | < 14 kg |
| Payload | 10 Gbit/s lasercom terminal for intersatellite links with GEO and downlinks to ground |
| Launch | 2023 (estimated) |
| GEO counterpart | Engineering Test Satellite 9 (ETS-9) |
| Operation | ≥ 6 months |

The GEO counterpart of CubeSOTA in the intersatellite links is the High-speed Communication with Advanced Laser Instrument (HICALI) onboard ETS-9 [8] [10] (Fig. 5). HICALI is a 10-Gbit/s bidirectional lasercom terminal operating in the 1550-nm range based on differential phase-shift keying (DPSK) modulation. It has a 15-cm telescope with a ±10º field of regard, enough to cover all the Earth disc with some additional margin. Its high-power amplifier (HPA) can transmit up to 2.5 W of optical power at 1540 nm and it has a receiver at 1560 nm with a low-noise amplifier (LNA) capable of detecting an optical power higher than 40 nW. The discrimination of uplink and downlink is implemented in 2 ways, i.e. by using different polarizations and wavelengths. The HICALI tracking system requires a 1530-nm beacon for the initial acquisition as well as to close the loop of the fine pointing system during the communication phase.

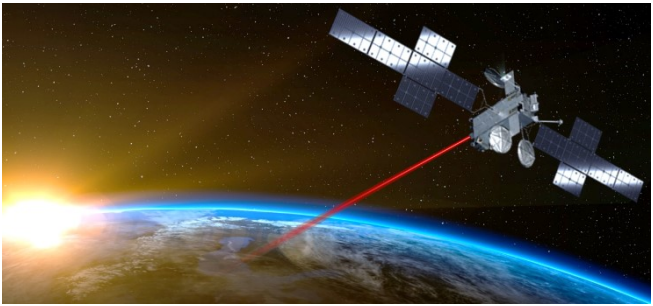

Fig. 5. Conceptual illustration of ETS-9 lasercom experiment.

At the time of the conception of the CubeSOTA mission, HICALI was already designed. Therefore, CubeSOTA had to adapt to the HICALI design with no possible modifications. Originally, HICALI was designed for GEO-ground bidirectional links only, not for intersatellite links. For example, the GEO relay application from LEO should be designed so that it is the GEO satellite the one transmitting the reference beacon to the LEO terminals, which are supposed to be small satellites with less onboard resources [11]. However, HICALI transmits only one fixed narrow communication beam, which cannot be used as a beacon in the LEO satellite. This requires implementing a high-power low-divergence beacon in the CubeSat so that HICALI's tracking system can close the loop and point towards the LEO counterpart. Furthermore, since the 1530-nm beacon must be maintained during the whole communication link, its divergence must be narrowed in order to allow to keep the beacon at a lower power and use most of the gain of the optical amplifier for the 1560-nm communication laser. A beam-divergence control will be implemented to allow this functionality.

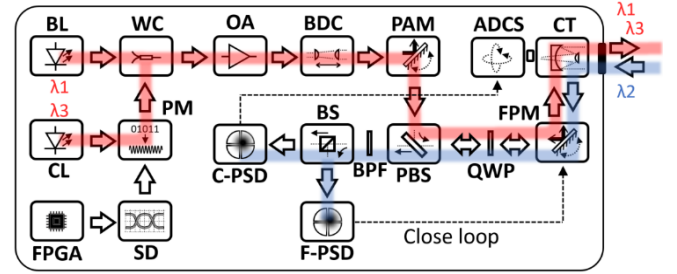

Fig. 6. CubeSOTA lasercom-system functional diagram

Figure 6 shows the functional diagram of CubeSOTA's lasercom system. The FPGA generates a 10-Gbit/s signal that, after the signal driver (SD), is used to modulate the communication laser (CL) with a phase modulator (PM), after which is combined with the beacon laser (BL) by a WDM coupler (WC). Both signals are amplified by an optical amplifier (OA) and their divergence is set by a beam-divergence control (BDC). The point-ahead mechanism (PAM) introduces an offset to the transmitted beams relative to the received one and is steered by a fine-pointing mechanism (FPM) before being transmitted through an 85-mm classical Cassegrain telescope (CT).

An isolation close to 100 dB must be achieved between the uplink and downlink beams. This will be implemented by discriminating both polarization and wavelength. Circular polarizations with opposite rotations are employed in the free-space and the received beam is separated from the transmitted one by a quarter-wave plate (QWP), to convert the received circular polarization into linear, followed by a polarized beam splitter (PBS) to separate both paths, and then by a spectral band-pass filter (BPF) for extra isolation. The received signal is detected in two position-sensitive detectors (PSD), one for coarse tracking and one for fine tracking.

Since the LEO-GEO intersatellite links are so technically challenging, the lasercom payload will be able to carry out experiments with the ground as well by using the NICT's 1-m OGS in Koganei, Tokyo. The required BDC, necessary to control the divergence of the transmitted beams, will be useful to explore the limits of the body pointing with the ADCS, as well as the fine pointing system with the FPM. Space lasercom systems are usually designed with a reasonable margin so that the link can be closed even under bad conditions. Being able to control the transmitted-beam divergence can allow making the most of the lasercom system.

Figure 7 shows a simulation of the LEO-GEO distance as well as the associated free-space losses during one week in 2023. It can be seen that roughly every half a day there is a repeating pattern of 8 approaches with 7 Earth eclipses in the middle. Between the maximum and minimum LEO-GEO distance, there is only a difference of 1.88 dB, which means that if there is line of sight, most of the time could be used for communications, with an average duration close to 1 hour, as shown in table 1. However, in the case of CubeSOTA, the scarce resources will not allow such long communication links. Specifically, the OA will probably not allow links longer than 5 minutes, thus this is the currently assumed duration for the experiments. The closest approaches were selected to carry out the experiments, which happen at ~35,500 km. However, even in the most-distant of the 8 approaches, the minimum distance is only 0.65 dB away from the closest distance of the closest approach, hence the minimum duration between two consecutive experiments could be only 99 minutes, provided that there is enough power in the CubeSat batteries.

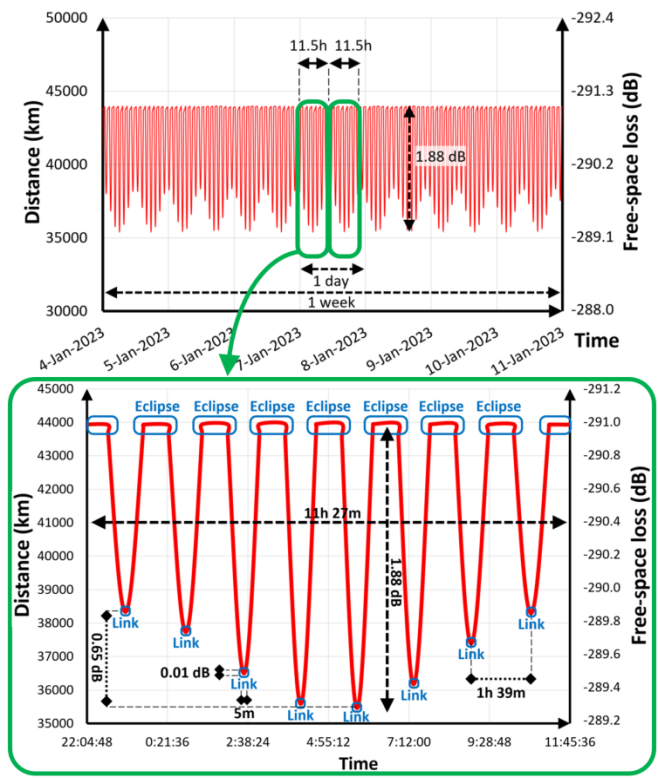

Fig. 7. LEO-GEO distance and free-space loss simulation.

To be able to close the 10-Gbit/s LEO-GEO links, the telescope divergence must be diffraction-limited. With such a narrow beam, the so-called point-ahead angle (PAA) needs to be considered, since it takes some time for the transmitted beam from the CubeSat to reach the GEO satellite due to the finite speed of light. Therefore, downlink and uplink beams are angularly separated by a PAA. Fig. 8 shows this phenomenon as well as the worst-case theoretical calculation between a LEO and a GEO satellite, which is ~53 µrad.

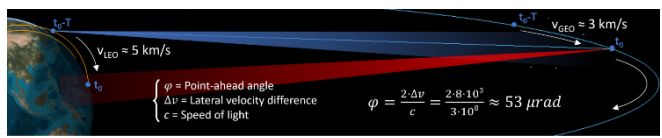

Fig. 8. LEO-GEO point-ahead angle calculation.

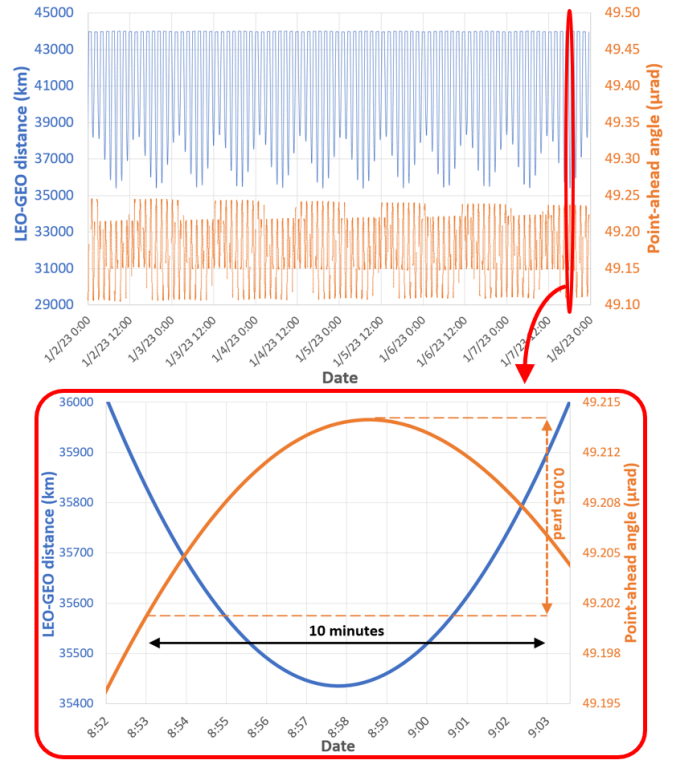

Fig. 9. LEO-GEO point-ahead angle simulation.

Figure 9 shows a simulation of the PAA during several days of the CubeSOTA mission. On the bottom, the central 10 minutes of the closest approach are zoomed in since at this point the PAA reaches its maximum value. The PAA variation is very small during the experiment time, thus it can be considered to be a fixed value of 49.2 µrad. This maximum PAA angle was confirmed to be achievable by the HICALI lasercom terminal. It was a key step in the feasibility study, since HICALI cannot be modified and it was designed for smaller PAA angles during the GEO-ground links.

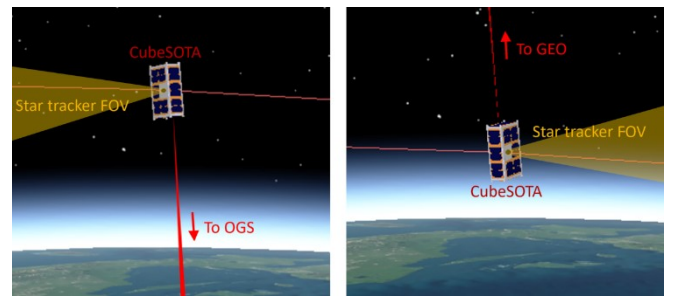

Fig. 10. Star-tracker to allow links with GEO and ground.

In order to be able to carry out experiments with both GEO and ground, the star tracker of the ADCS system was placed at a 90º-angle from the lasercom telescope, as shown in Fig. 10. In a more conventional LEO-GEO intersatellite link, the star tracker of the LEO terminal would be placed in the same direction of the telescope (with some offset to avoid the link from the GEO satellite), which makes sure that there will be stars always within its field of view (FOV).

Different mission-success criteria were defined for the CubeSOTA mission (table 3), being the minimum success the acquisition of the beacon signal in HICALI for the LEO-GEO intersatellite links, in the OGS for the downlinks with the ground, as well as the acquisition of the ground beacon in the

CubeSat. Full success will be achieved when the downlink signal from HICALI is acquired in the CubeSat, as well as when 10-Gbit/s communication is carried out with the OGS. Finally, the extra success consists of the 10-Gbit/s communication link from LEO to GEO, given the challenging requirements to be able to close the link in this case.

Table 3. Levels of success criteria in CubeSOTA mission.

| Minimum success | • Acquisition of CubeSat beacon in GEO<br>• Acquisition of CubeSat beacon in OGS<br>• Acquisition of OGS beacon in CubeSat |
|---|---|
| Full success | • Acquisition of GEO downlink in CubeSat<br>• 10-Gbit/s communication with OGS |
| Extra success | • 10-Gbit/s communication with GEO |

The necessary sequence to achieve extra success makes use of all the CubeSOTA subsystems, and it is described as follows: First, the LEO terminal transmits a fixed high-power wide-divergence beacon towards its GEO counterpart. If HICALI cannot detect it, a spiral scan with a narrow beacon can be performed to increase the power density reaching the GEO satellite. When HICALI has detected the beacon, it automatically starts transmitting the 10-Gbit/s downlink signal, which is used in the CubeSat as a beacon (no communication receiver is implemented due to the scarce onboard resources) to close the body-pointing loop, as well as the fine-pointing loop. When this stage is achieved, CubeSOTA starts optimizing its beacon, by reducing its divergence until the minimum possible, and then by reducing the gain of the beacon wavelength. Then the 10-Gbit/s communication uplink beam is transmitted by using most of the OA's gain, while transmitting the beacon at a low optical power.

## IV. Conclusion

This paper described the current plans of NICT and the University of Tokyo to carry out a LEO-GEO intersatellite 10-Gbit/s lasercom demonstration mission by using a 6U CubeSat in LEO and the HICALI lasercom terminal onboard the ETS-9 GEO satellite. The basic design of the CubeSOTA lasercom system was explained, as well as the experiments this mission will consist of.